\documentclass[final,5p,times,twocolumn,authoryear]{elsarticle}
\pdfoutput=1
\usepackage{bbm}
\usepackage{mathrsfs}
\usepackage{slashed}
\usepackage{caption}
\usepackage{epstopdf}
\usepackage[normalem]{ulem}
\usepackage[bottom]{footmisc}
\usepackage{subcaption}
\usepackage{bbold} 
\usepackage{titlesec}
\usepackage{threeparttable}
\usepackage{booktabs}
\usepackage{changepage}
\usepackage[utf8]{inputenc}
\usepackage{dsfont} 
\usepackage{grffile}
\usepackage{graphicx}  
\usepackage{dcolumn}   
\usepackage{bm}        
\usepackage{amssymb}   
\usepackage{setspace}
\usepackage{amsmath, amssymb, setspace}
\usepackage{array}
\usepackage{booktabs}
\usepackage{caption}
\usepackage{indentfirst}
\usepackage{float}
\usepackage{lmodern}
\usepackage{multirow}
\usepackage{soul}
\usepackage[normalem]{ulem}
\usepackage{braket}
\usepackage{comment}
\usepackage[draft]{pgf}
\usepackage{adjustbox} 
\usepackage{xspace} 

\newcommand{\DoBox}[1]{\begin{center}
\color{red}\fbox{
\begin{minipage}{0.9\textwidth}
\end{minipage}}
\end{center}}

\journal{Physics Letters B}

\usepackage[most]{tcolorbox}
\usepackage{empheq}

\begin{document}
\begin{frontmatter}

\title{Constraining Axion-Like Particles Dark Matter in Coma Berenices with FAST}

\author[label1,label2]{Wen-Qing Guo}
\author[label1]{Zi-Qing Xia$^*$}
\ead{Corresponding author:xiazq@pmo.ac.cn}
\author[label1,label2]{Xiaoyuan Huang$^*$}
\ead{Corresponding author:xyhuang@pmo.ac.cn}
\affiliation[label1]{organization={Key Laboratory of Dark Matter and Space Astronomy, Purple Mountain Observatory, Chinese Academy of Sciences},
             city={Nanjing},
             postcode={210033},
             country={China}}

\affiliation[label2]{organization={School of Astronomy and Space Science, University of Science and Technology of China},
             city={Hefei, Anhui},
             postcode={230026},
             country={China}}

\begin{abstract}
Axions and axion-like particles (ALPs) appear in many extensions of the Standard Model and are being investigated as promising dark matter (DM) candidates. 
One viable methodology for their detection involves the investigation of the line-like radio emissions from the dwarf spheroidal galaxy, potentially originating from the radiative decay of ALPs or the conversion of ALPs in the magnetic field. 
In this work, we constrain the properties of ALPs using the 2-hour radio observation of Coma Berenices through the Five-hundred-meter Aperture Spherical radio Telescope (FAST).
The $\rm 95\%$ upper limits of the ALP-photon coupling are calculated for the ALP decay and conversion scenarios, respectively.
Note that the sensitive ALP masses for FAST range from $\sim \mu \rm eV$ to tens of $\mu \rm eV$, where ALP can explain the DM abundance naturally.
However, our limits are weaker than those of the CAST helioscope, which can provide an independent and complementary check on the ALP non-detection for ground experiments. Furthermore, we evaluate
the expected sensitivity on the ALP
of FAST with its full designed bandwidth (70 $\rm MHz$ - 3 $\rm GHz$) for 100 hours of observation time. Our results indicate that, even with the exceptional sensitivity of the FAST, it is challenging to surpass the existing experimental constraints on ALP DM using radio observation of dSphs, unless the possible enhancements of ALP signals by compact stars in dSphs are considered. 
\end{abstract}

\end{frontmatter}

\section{Introduction}

Based on a lot of astronomical and cosmological observations in recent years, the existence of dark matter (DM) has become widely accepted. Notably, studies of the cosmic microwave background (CMB) within the framework of the standard cosmological model indicate that DM contributes approximately a quarter of the total energy density of the universe~\citep{Planck:2018vyg}. Unraveling the nature of dark matter has become a pivotal pursuit in modern cosmology, given its fundamental role in the evolution of the universe.

Dark matter is hypothesized to be elementary particles beyond the Standard Model (SM). These hypothetical particles should be neutral, have very weak interactions with SM particles, and exhibit remarkable stability over the cosmic time scales. In addition, DM is required to be sufficiently cold to allow for efficient structure formation in the universe~\citep{Blumenthal:1984bp}. In light of these considerations, weakly interacting slim particles (WISPs) with sufficiently small masses can provide a scenario to realize all these features~\citep{Ringwald:2012hr, Arias:2012az}. Axion-like particles (ALPs), in particular, stand out as highly promising candidates within the WISP paradigm, finding inspiration in string theory as extensions of the SM~\citep{Irastorza:2018dyq}.

Apart from thermally produced as light DM, ALPs can also be non-thermally produced through the misalignment mechanism in the early universe, covering a broad mass range of $10^{-20}\,\rm eV \leq m_{a} \leq eV$~\citep{Preskill:1982cy, Abbott:1982af, Dine:1982ah, Adams:2022pbo}. These ALPs could naturally persist as a population of cold dark matter until the present day~\citep{Arias:2012az, Marsh:2015xka, Irastorza:2018dyq, Co:2020xlh, AyadMohamedAli:2021unr}. While ALPs share similar properties with QCD axions,
ALPs are generally unrelated to the Peccei-Quinn (PQ) mechanism~\citep{Irastorza:2021tdu}, without a dependence between the coupling and the mass. The coupling of ALPs to photons allows for their conversion into photons through the Primakoff effect in the presence of an external magnetic field, and their radiative decay into pairs of photons. These processes form the theoretical foundation for the search for ALPs in laboratory experiments~\citep{ADMX:2001dbg, Bahre:2013ywa, ADMX:2019uok} and from astrophysical observations~\citep{Fermi-LAT:2016nkz, CAST:2017uph, 2018PhRvD..97f3009Z,2018PhRvD..97f3003X,2019JCAP...06..042L,2019PhRvD.100l3004X,Li:2020pcn}.  
Particularly, for ALP masses around $\mu \mathrm{eV}$, photons within very narrow energy, generated from ALPs conversion in the magnetic field~\citep{Hook:2018iia, Wang:2021wae, Foster:2022fxn,Sigl:2017sew, Caputo:2018ljp} or ALP decay~\citep{Caputo:2018ljp, Caputo:2018vmy},
would fall within the frequency coverage of radio telescopes. Thus, radio observations of DM-rich systems, such as dwarf spheroidal galaxies (dSphs), offer a promising avenue for searching for ALPs and further constraining their properties. This approach could serve as a valuable complement to ground-based ALP detection experiments, such as the CAST helioscope~\citep{CAST:2017uph} and the ADMX haloscope~\citep{ADMX:2001dbg, ADMX:2019uok}.

In this work, we perform radio searches for ALPs utilizing the Five-hundred-meter Aperture Spherical radio Telescope (FAST), which is the largest filled-aperture, single-dish, radio antenna in the world and has a very high sensitivity~\citep{2020RAA....20...64J, 2020Innov...100053Q}. FAST is located at a geographic latitude of $25^{\circ}29'10''$ and can observe a large portion of the sky, with a maximum zenith angle of $40^{\circ}$. Currently, the 19-beam receiver equipped on the FAST covers the frequency range of 1-1.5 GHz, corresponding to the $\mu {\rm eV}$ ALPs. 
In this paper, we selected the Coma Berenices dwarf galaxy as our target to constrain the properties of ALP, considering its observability with FAST and its proximity to the Earth.

The organization of this paper is as follows. 
In Section~\ref{Target and Expected Signal}, we introduce the properties for the Coma Berenices and calculate the expected signal from ALPs. 
In Section~\ref{Data Analysis and Results}, we conduct the FAST data analysis, place the upper limits on ALPs coupling parameter, and further discuss the results. 
Then we conclude in Section~\ref{Conclusion}.
Note that we use natural units and Lorentz-Heaviside units in our paper, which are $k_{B}=c=\hbar=1$ and $\epsilon_{0}=\mu_{0}=1$ respectively.

\section{Target and Expected Signal}\label{Target and Expected Signal}

\subsection{Coma Berenices}

Dwarf spheroidal galaxies (dSphs) of the Local group are considered among the most promising targets to detect DM signals, owing to their high mass-to-light ratios~\citep{Strigari:2007at, Fermi-LAT:2010cni}, and their short distances to us. dSphs are also generally believed to be free of radio background since they lack gas, dust, and stellar activity. Coma Berenices, which was discovered in the Sloan Digital Sky Survey~\citep{SDSS:2006fpg}, located at a distance of about 44 kpc from the Sun, centered at $\rm RA = 12h26m59s$, $\rm DEC = +23^{\circ} 54'00''$~\citep{Kalashev:2020hqc} in equatorial coordinates (J2000), is one of the darkest galaxies known so far~\citep{Simon:2007dq, Battaglia:2022dii} and is expected to have a strong DM signal~\citep{Bonnivard:2015xpq}. Coma Berenices also experiences negligible tidal disruption from the Milky Way~\citep{Munoz:2009hj}, implying that its stellar kinematic characteristics are primarily dominated by the gravitational potential of its dark matter halo. In addition, the location of Coma Berenices falls within the proper zenith angles suitable for FAST. Thus, we use the FAST observation of Coma Berenices to search for the ALP signals in this work.

The expected fluxes of ALP signals depend on the spatial distribution of ALP within the dark matter halo. Following the work analyzing stellar kinematic data~\citep{Bonnivard:2015xpq}, we can model the dark halo profile of Coma Berenices as the Einasto profile 
\begin{equation}
    \rho(r)=\rho_{0}\exp\left[-\frac{2}{\alpha}\left(\left(\frac{r}{r_{s}}\right)^{\alpha} - 1\right)\right],
\end{equation}
where $\alpha = 0.847$, $r_{s} = 5.14\,{\rm kpc}$, and $\rho_{0} = 1.962\,{\rm GeV\,cm^{-3}}$ represent the best-fitting values of logarithmic slope, scale radius and normalization, respectively.

\subsection{ALPs Decay}
Whether in ground-based experiments or astrophysical observations, the predominant methods for searching for ALPs mostly rely on the coupling between ALPs and photons.
The basic process of the interaction between ALP and photons is the ALP-two-photons vertex, and the Lagrangian equation can be written as
\begin{equation}
    L_{a\gamma}=-\frac{1}{4}g_{a\gamma}a F_{\mu \nu} \Tilde{F}^{\mu \nu},
\end{equation}
where $a$ is the ALP field, $F_{\mu \nu}$ is the electromagnetic field strength, $\Tilde{F}^{\mu \nu}$ is its dual, and $g_{a\gamma}$ is the ALP-photon coupling parameter. Therefore, the coupling between ALPs and photons would result in the decay of ALPs into two photons and the conversion between ALPs and photons through the Primakoff effect in the presence of an external electromagnetic field.

In a vacuum, the spontaneous decay of an ALP with the mass of $m_{a}$ would generate two photons, each with a frequency $\nu_{0} = m_{a}/4\pi$, and the lifetime could be expressed in terms of ALP mass and ALP-photon coupling by
\begin{equation}\label{decay time}
    \tau_{a} = \frac{64\pi}{m^{3}_{a}g_{a\gamma}^{2}}.
\end{equation}

The standard flux density from the spontaneous decay of ALPs in one dwarf galaxy is then given by
\begin{equation}\label{spontanoues decay0}
    S_{{\rm ALP},spon.} = \frac{1}{4\pi \Delta\nu\tau_{a}}\int d\Omega dl\rho(l,\Omega),
\end{equation}
where $\rho(l,\Omega)$ is the ALP density at distance $l$ along the line of sight and in the solid angle $\Omega$, and $\Delta \nu$ is the width of the ALP line determined by the ALPs velocity dispersion $\sigma_{\mathrm{disp}}$, with $\Delta \nu = \nu_{0}\sigma_{\mathrm{disp}}$. 
In this work, we take ALPs velocity dispersion to be $\sigma_{\rm disp}=4\, \rm km s^{-1}$~\citep{Bonnivard:2015xpq,Caputo:2018vmy, Battaglia:2022dii} in Coma Berenices. 
Furthermore, with the definition of the $D$ factor~\citep{Bonnivard:2015xpq}, we could rewrite Eq. ~(\ref{spontanoues decay0}) as 
\begin{equation}\label{spontaneous decay}
    S_{{\rm ALP},spon.} = \frac{m^{2}_{a} g_{a\gamma}^{2}}{64\pi} \frac{D(\alpha_{\rm int})}{\sigma_{\rm disp}},
\end{equation}
where $\alpha_{\rm int}$ is the integrating angle, corresponding to the angular size of the region of interest (RoI). In this work, we calculate the flux densities within the RoI with $\alpha_{\rm int} =3'$, which is the FWHM of the FAST central beam (Beam 01)~\citep{2020RAA....20...64J} .

Due to the very long lifetime of ALPs, detecting signals from their spontaneous decay poses a challenge for current radio telescopes. However, signals could also be enhanced via stimulated decay while the ALP decay happens in an ambient radiation field~\citep{Caputo:2018vmy}.  Quantitatively, the stimulated emission can be described by multiplying the spontaneous emission rate by a factor of $2\,f_{\gamma}$~\citep{Caputo:2018vmy}. In the case of the $\sim \mu \rm eV$ ALPs, the stimulated decay would take place in radiation fields at radio frequencies, which are made up of the CMB radiation, Galactic diffuse emission and extragalactic radio background. The stimulated emission will follow not only the profile of ALPs but also the profile of the ambient photon field. As Coma Berenices with high galactic latitude and low  Galactic diffuse radio emission, the efficient photon background would only consists of the CMB and the extragalactic radio background. Both of them are followed by the blackbody spectra and the isotropic spatial distribution. The photon occupation number from the blackbody spectrum can be given as~\citep{Caputo:2018vmy}
\begin{equation}
    f_{\gamma}=\frac{1}{e^{E_{\gamma}/T}-1},
\end{equation}
where the photon energy is $E_{\gamma}=m_{a}/2$. For the CMB, the corresponding blackbody temperature is $T_{\rm CMB}=2.75\, \rm K$. While for the extragalactic radio background, a frequency-dependent temperature is given by $T_{\rm ext}(\nu) \approx 1.19({\rm GHz}/{\nu})^{2.62}\, \rm K$~\citep{Fixsen:2009xn, Fornengo:2014mna, Caputo:2018vmy}. Thus, the total flux density from stimulated decay can be written as
\begin{equation}\label{stimulated decay}
    S_{{\rm ALP},stim.}=2\frac{m^{2}_{a}g_{a\gamma}^{2}}{64\pi} \frac{D(\alpha_{\rm int})}{\sigma_{\rm disp}} (f_{\gamma,{\rm CMB}}+f_{\gamma,{\rm ext}}).
\end{equation}
The signals originate from the decay of ALPs, involving both spontaneous and stimulated decay processes. Consequently, the total flux density originating from the decay mechanism can be expressed as the sum of { Eq.~(\ref{spontaneous decay}) and Eq.~(\ref{stimulated decay})}.

\subsection{ALPs Conversion}
 
Apart from the decay process, the Primakoff effect would allow the ALP to convert to a photon with energy $E_{\gamma}=m_{a}$ in an external magnetic field. 
The photon flux arising from ALPs conversion is proportional to the square of the magnetic field strength $B_{\rm rms}^{2}$ and a suppression factor $f(m_{a})\simeq (m_{a}l_{c})^{n}$, where $n$ is $-2/3$ for the Kolmogorov turbulence and $l_{c}$ is the magnetic field coherence length~\citep{Sigl:2017sew, Caputo:2018ljp}. The exact strength of the magnetic field in dwarf galaxy remains unknown. Here we take the magnetic field strength in Coma Berenices as $B_{\rm rms}=1\, \rm \mu G$, which is assumed as a typical magnetic field strength in dSphs~\citep{McDaniel:2017ppt, Kar:2022ngx}. To estimate the magnetic field coherence length, we adopt the value of $l_{c}=0.01\,\mathrm{pc}$~\citep{Kar:2022ngx} with an assumption that the coherence length of the magnetic field in dSphs is significantly smaller than that in the Galactic center~\citep{Caputo:2018ljp}.


Thus, for dwarf galaxies, the flux density from the ALP-photon conversion could be given as ~\citep{Caputo:2018ljp}
\begin{equation}
    S_{{\rm ALP},conv.}=\frac{B^{2}_{\rm rms} g_{a\gamma}^{2}}{m^{2}_{a}\sigma_{\rm disp}}f(m_{a})D(\alpha_{\rm int}),
\end{equation}
where the integrating angle $\alpha_{\rm int} =3'$ is the same as that of the ALPs decay.

\begin{figure*}[htp]
\centerline{
\includegraphics[width=1\columnwidth]{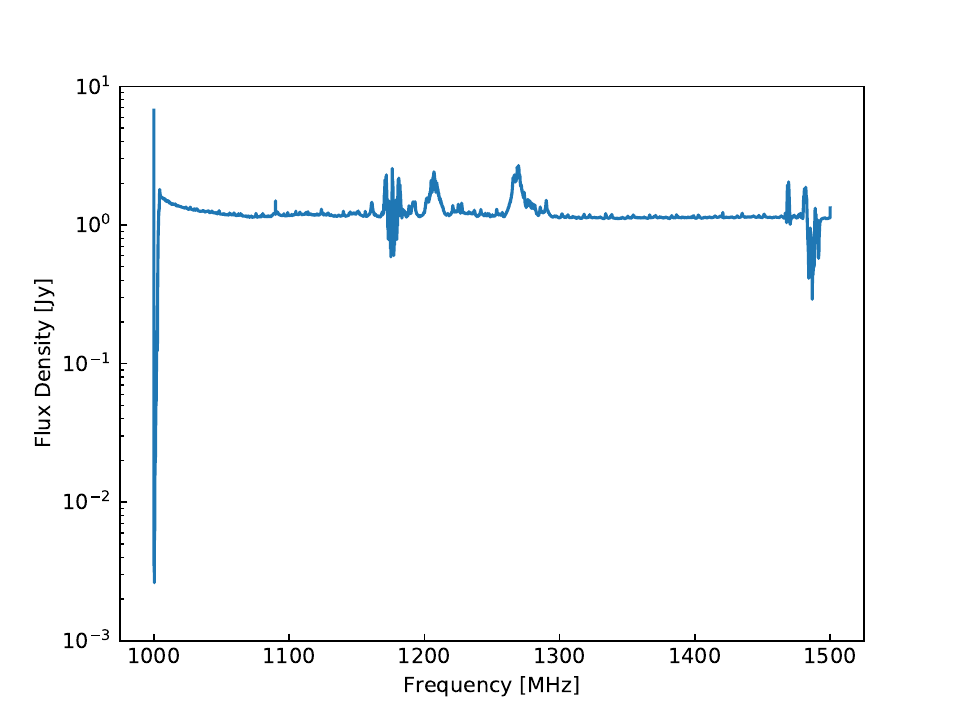}
\includegraphics[width=1\columnwidth]{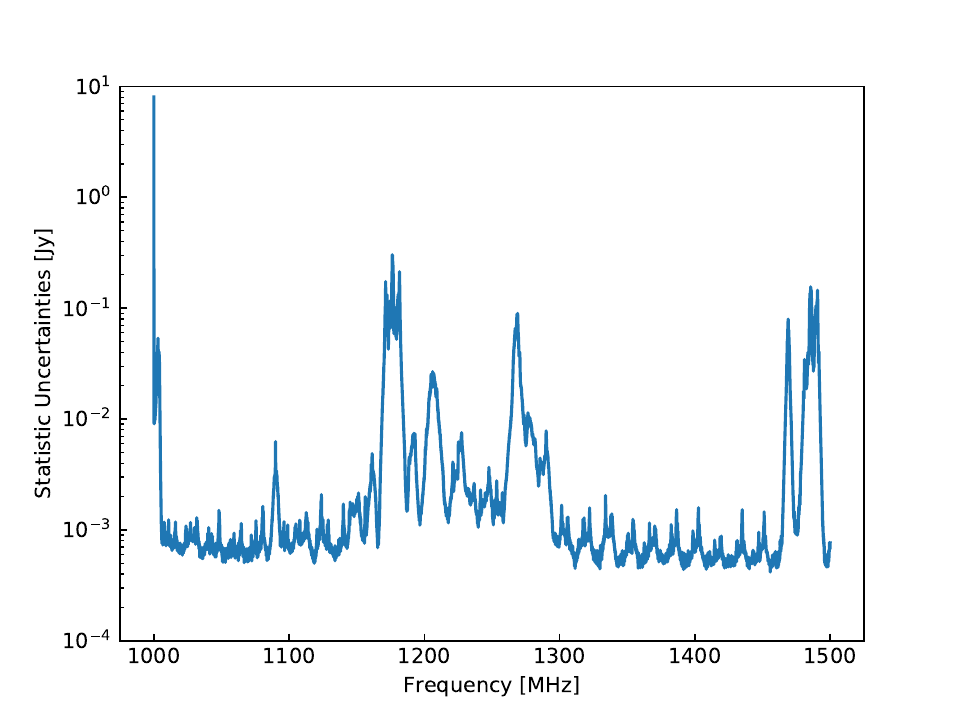}}
    \caption{{\it Left panel} - The time-averaged flux density $\bar D$ of the ``ON source" from FAST observation in the frequency range $1,000-1,500\,\mathrm{MHz}$. {\it Right panel} - The statistical uncertainties $\sigma$ as a function of frequency for the 2-hour FAST observation. }
    \label{fig:flux_density}
    
\end{figure*}

\section{Data Analysis and Results}\label{Data Analysis and Results}

\subsection{Observation}
{The FAST observation was conducted} from 07:00:00 to 08:50:00 on December 14, 2020. Utilizing 19-beam receiver, the data were recorded in two {orthogonal polarizations, namely XX and YY,} across 65,536 spectral channels, encompassing the frequency range of $1,000$-$1,500$ MHz. The frequency resolution achieved was $\mathcal{B}_{\rm res}=7.63\, \rm kHz$. In our observation, we utilized the SPEC(W+N) backend with a sampling time of 1 second. Determined by our original motivation to constrain the WIMP property~\citep{Guo:2022rqq}, the ``ON/OFF" { observation mode of FAST} was chosen to reduce the contamination from the earth's atmosphere, baseline variation, and other foregrounds. During each observation round, the central beam of the FAST is directed towards the Coma Berenices for a duration of 350 seconds. Following this, the central beam is shifted to the ``OFF source" position, located half a degree away from the ``ON source" and devoid of known radio sources, for an additional 350 seconds. The total observation time is two hours, comprising 9 rounds of ``ON/OFF" observation. We utilized the low-noise injection mode to calibrate the signal, featuring a characteristic noise temperature of approximately 1.1 K~\citep{2020RAA....20...64J}. Throughout the entire observation, the noise diode underwent a continuous on-and-off switching cycle, each lasting for one second. In this work, we are going to utilize the ``ON source" data only, but use an alternative method to estimate the background as in Ref.~\citep{An:2022hhb}, instead of using the ``OFF source" data.

\subsection{FAST data analysis}

The original instrument's recordings, when taken without and with noise injection, are denoted by $P^{\rm cal_{off}}$ and $P^{\rm cal_{on}}$ respectively. Subsequently, the antenna temperature $T_{\rm A}$ can be calibrated as~\citep{ONeil:2002amr,2020RAA....20...64J,2021MNRAS.503.5385Z}
\begin{equation}\label{Temperature calibration}
    T_{\rm A}=\frac{P^{\rm cal_{off}}}{P^{\rm cal_{on}}-P^{\rm cal_{off}}}T_{\rm noise},
\end{equation}
where $T_{\rm noise}$ is the noise temperature determined through the measurements with hot loads. {Thus, after temperature calibration, the data is organized as follows: \{{\it time}, {\it frequency}, {\it polarization} ({{$T_{\rm XX}$}} \& {$T_{\rm YY}$})\}. We followed the guidelines for FAST data analysis to verify the consistency of $T_{\rm XX}$ and $T_{\rm YY}$. In each frequency, we calculate $T_{\rm YY} - T_{\rm XX}$ for each time bin and obtain the standard deviation $\sigma_{(T_{\rm YY} - T_{\rm XX})}$. Time bins where $T_{\rm YY} - T_{\rm XX}$ exceeds $5\sigma_{(T_{\rm YY} - T_{\rm XX})}$ from zero are masked, as done in our previous work \cite{Guo:2022rqq, An:2022hhb}. However, $T_{\rm YY} - T_{\rm XX}$ in all time bins of our data is consistent with zero, and no data are masked here. The two polarizations are then combined into the total intensity, i.e., $T=(T_{\rm XX}+T_{\rm YY})/2$.} Following these steps, we derive the antenna temperature at the ON position for each time and frequency bin of the central beam, denoted as $T_{\rm ON}$. Additionally, the time bins during which the telescope's stability is affected by the transition between ON and OFF positions are masked. This process results in the retention of 1116 time bins for the ON observations of each frequency \footnote{We discarded the last three rounds of observations in our previous work~\citep{Guo:2022rqq} because the baseline of the OFF observations was unstable.}. We then utilize the pre-measured antenna gain, which depends on both beam and frequency~\citep{2020RAA....20...64J}, to convert the temperature $T_{\rm ON}$ into flux density. 

{
In each time bin, we conduct a 2-fold down-binning analysis in frequency to attain a resolution of $15.26\,\mathrm{kHz}$. Subsequently, the number of spectral channels is condensed to 32768.
Thus, the expected line-like signal can be completely contained in seven adjacent channels.
Besides, the data may suffer from short-period artificial radio frequency interference (RFI) during the data recording process, leading to significant fluctuations in short time series.
To ensure the good quality of data, we adopt a data cleaning process to select good data in the time domain, as performed in Ref.~\citep{An:2022hhb}. 
That is, for each frequency bin $i$, we divide the data from 1116 time bins into 34 groups. The first 33 groups contain 33 time bins and the last group contains 27 time bins. We select the group with the smallest variance, $\sigma^{2}_{i, {\rm ref}}$, as the reference group. The mean value of the reference group is denoted as $\mu_{i, {\rm ref}}$. We then retain the good data with deviation from the reference mean smaller than $5\sigma_{i, {\rm ref}}$, i.e. $|D_{i}-\mu_{i, {\rm ref}}|<5\sigma_{i, {\rm ref}}$, where $D_{i}$ is the data along the time bins. Finally, we compute the time-averaged flux density $\bar{D}_i$ by averaging across time bins for each frequency bin $i$. The corresponding statistical uncertainty $\sigma_i$ is defined as the standard deviation of $D_{i}$.
Accordingly, the time-averaged flux density and statistical uncertainty  for all frequencies, obtained after the calibration and data-cleaning processes, are shown in the left and right panels of Fig.~\ref{fig:flux_density}, respectively.}

\begin{figure*}[htp]
\centerline{
\includegraphics[width=1\columnwidth]{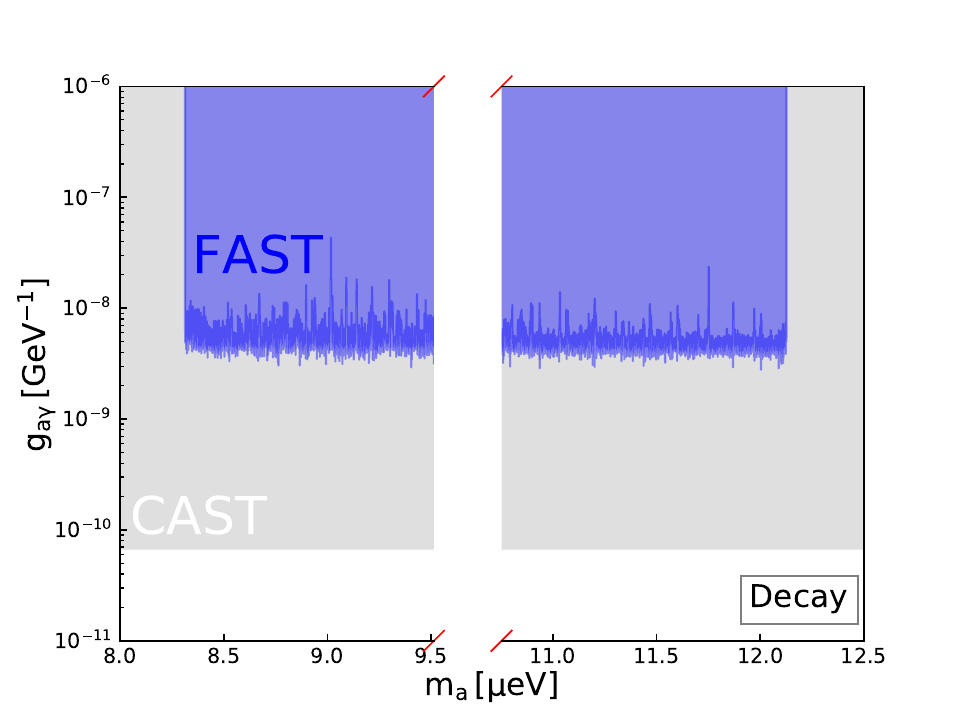}
\includegraphics[width=1\columnwidth]{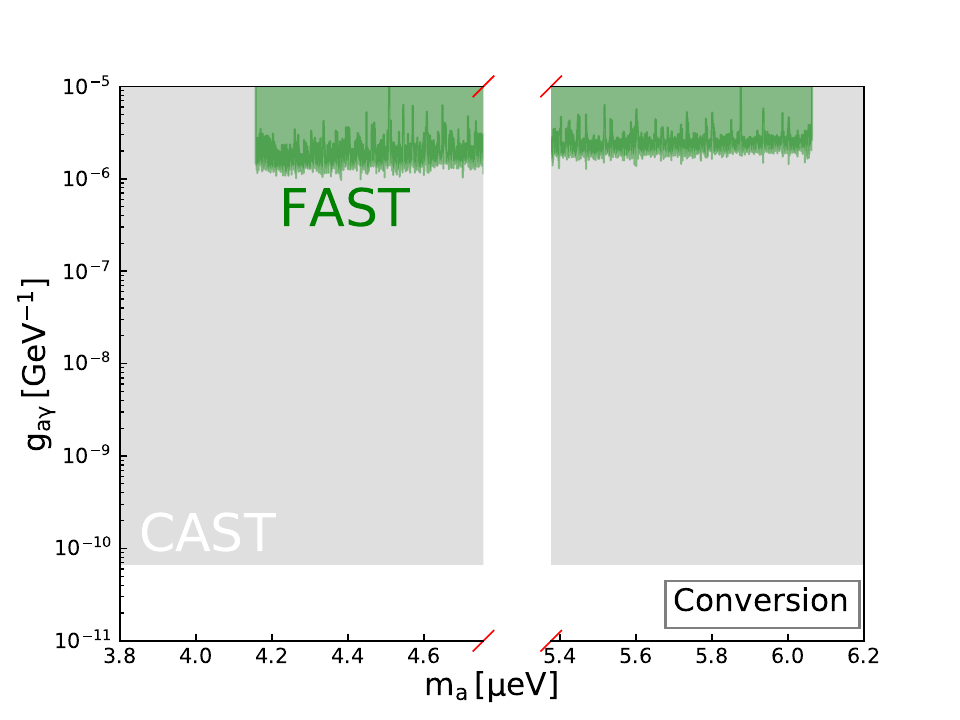}}
    \caption{{\it Left panel} - The 95\% upper limits of $g_{a\gamma}$ in the ALP mass range [8.30, 12.12] $\mu \rm eV$ based on the FAST observation of Coma Berenices (solid blue) for the ALPs decay process. {\it Right panel} - The 95\% upper limits of $g_{a\gamma}$ in the ALP mass range [4.15, 6.06] $\mu \rm eV$ using the same FAST data (solid green) for the ALPs conversion process in the galactic magnetic field. In {\it both} panels, the results are compared with the existing constraints from CAST~\citep{CAST:2017uph}.}
    \label{fig:axion}
\end{figure*}

Assuming a Gaussian shape of the ALP signal, the mean flux density in channel $i$ is then given by
\begin{equation}
\begin{split}
    \bar{S}_{i}(m_{a}, g_{a\gamma})=&S_{{\rm ALP}, n}(m_{a}, g_{a\gamma}) \\
    &\times \int_{\nu_{i, \mathrm{min}}}^{\nu_{i, \mathrm{max}}}\frac{d\nu}{\sqrt{2\pi}\Delta \nu} \mathrm{e}^{{-(\nu-\nu_{0})^{2}}/{2(\Delta \nu)^{2}}},
\end{split}
\end{equation}
where $\nu_{i, {\rm min}}$ and $\nu_{i, {\rm max}}$ are the start and end frequencies of the $i$-th channel. And $\nu_{0}$ is equal to $m_{a}/4\pi$ or $m_{a}/2\pi$ when $n$ represents the decay or conversion mechanism, respectively. { $\Delta \nu = \nu_{0}\sigma_{disp}$ is the width of the ALP line as defined in Eq.~(\ref{spontanoues decay0})}. 

For one selected frequency bin $i_{0}$, we model the local background in the bin set $\mathcal{F}(i_{0}-k, i_{0}+k)$, i.e. the bins around the $i_{0}$ frequency bin from the $i=i_{0}-k$ bin to the $i=i_{0}+k$ bin, with a polynomial function~\citep{An:2022hhb},
\begin{equation}\label{background definition}
    B(a, \nu)=a_{n}\nu^{n}+a_{n-1}\nu^{n-1}+...+a_{1}\nu+a_{0},
\end{equation}
where we choose $k=10$ and $n=3$. {Thus, we treat $a=(a_{3}, a_{2}, a_{1}, a_{0})$ as nuisance parameters in the subsequent likelihood analysis.}   We fit the background by minimizing the function
\begin{equation}\label{background}
    \sum_{i=i_{0}-k}^{i_{0}-b}\frac{(B(a,\nu_{i})-\bar{D}_{i})^{2}}{\sigma^{2}_{i}} + \sum_{i=i_{0}+b}^{i_{0}+k}\frac{(B(a,\nu_{i})-\bar{D}_{i})^{2}}{\sigma^{2}_{i}},
\end{equation}
where $b$ is taken as 3 to avoid possible contamination from the ALP signal, considering the width of the signal. The result is labeled as $\Tilde{a}$ so that the background $B(\Tilde{a}, \nu)$ minimizes Eq.~(\ref{background}). Then the systematic uncertainty in this bin set can be estimated by the deviation of the data to the fitted background curve defined as
\begin{equation}
    (\sigma^{\rm sys})^{2}=\frac{1}{2k-2b-1}\left(\sum_{i=i_{0}-k}^{i_{0}-b}(\delta_{i}-\Bar{\delta})^2 + \sum_{i=i_{0}+b}^{i_{0}+k}(\delta_{i}-\Bar{\delta})^2\right),
\end{equation}
where
\begin{equation}
    \delta_{i}=B(\Tilde{a}, \nu_{i})-\bar{D}_{i}.
\end{equation}
We then estimate the total uncertainty of bin $i$ in the bin set $\mathcal{F}(i_{0}-k, i_{0}+k)$ as a sum of statistical uncertainty and systematic uncertainty, 

\begin{equation}
    (\sigma^{\rm tot}_{i})^{2} = \sigma_{i}^{2}+(\sigma^{\rm sys})^{2}.
\end{equation}

To constrain the property of ALPs, the least squares function for ALP DM with mass $m_{a}$ and coupling $g_{a\gamma}$ can be written as,
\begin{equation}
    \chi^{2}(m_{a}, g_{a\gamma}, a)=\sum_{i=i_{0}-k}^{i_{0}+k}\left (\frac{B(a,\nu_{i})+\bar{S}_{i}(m_{a}, g_{a\gamma})-\bar{D}_{i}}{\sigma^{\rm tot}_{i}}\right)^{2}.
\end{equation}

For a given ALP mass, we determine the 95\% confidence level upper limit of the coupling parameter $g_{a\gamma}$, denoted as $g_{a\gamma}^{\mathrm{lim}}$, by requiring $\Delta \chi^{2}(g_{a\gamma}^{\mathrm{lim}})=\chi^{2}(g_{a\gamma}^{\mathrm{lim}}, \hat{\hat{a}})-\chi^{2}(\hat{g}_{a\gamma}, \hat{a})=2.71$, where $\hat{g}_{a\gamma}$ and $\hat{a}$ represent the best-fit values of the coupling and nuisance parameters, respectively, obtained when the minimum $\chi^{2}_{\mathrm{min}}$ is achieved, and $\hat{\hat{a}}$ is the best-fit value of nuisance parameters when varying $g_{a\gamma}$. We repeat the above procedure to find $g_{a\gamma}^{\mathrm{lim}}$ over all the frequency bins .

\begin{figure*}[htp]
\centerline{
\includegraphics[width=1\columnwidth]{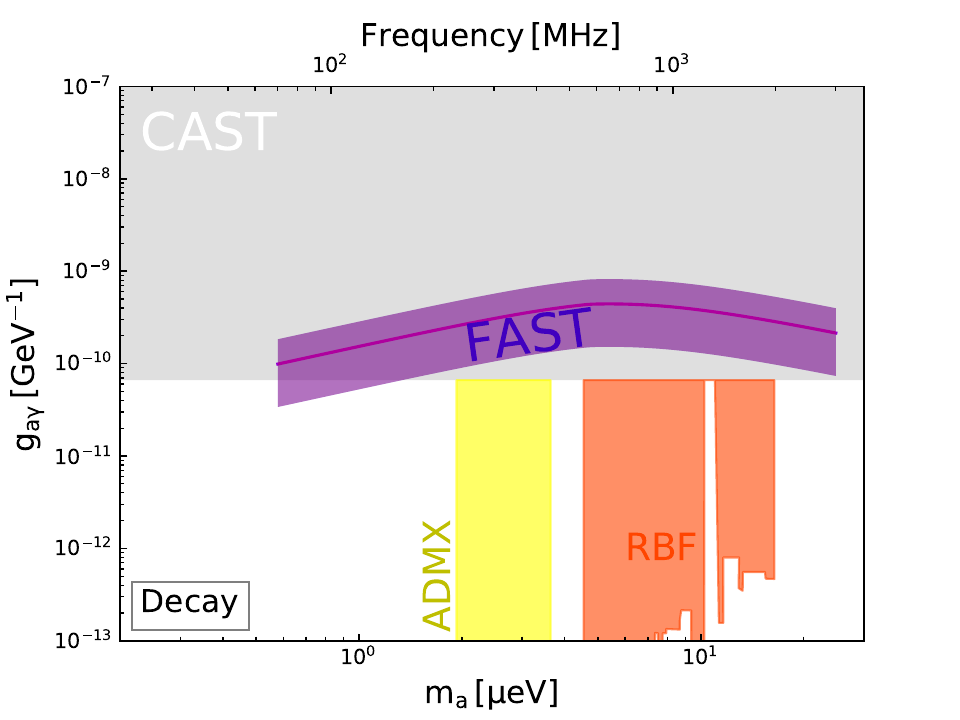}
\includegraphics[width=1\columnwidth]{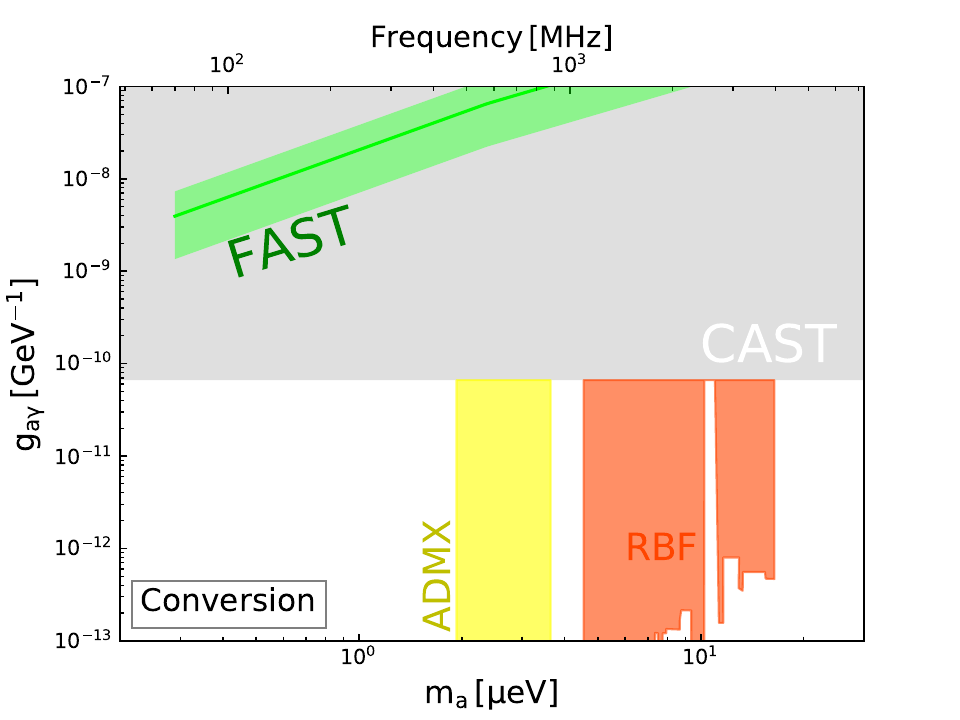}}
\caption{{\it Left panel} - Sensitivity of FAST to ALP mass range [0.58, 24.82] $\mu \rm eV$. The solid purple line (bands) represents the median ($\pm 95\%$ confidence intervals) sensitivity to ALP decay in Coma Berenices given the uncertainty of $D(\alpha_{int})$. {\it Right panel} - Sensitivity of FAST to ALP mass range [0.29, 12.41] $\mu \rm eV$. The solid green line (bands) represents the median ($\pm 95\%$ confidence intervals) sensitivity to ALP conversion in the presence of the galactic magnetic field of Coma Berenices given the uncertainty of $D(\alpha_{int})$. 
In {\it both} panels, sensitivity estimates are compared with the existing constraints from the ADMX haloscope~\citep{ADMX:2001dbg, ADMX:2019uok} and the CAST helioscope ~\citep{CAST:2017uph}.}
\label{fig:forecast}
\end{figure*}

\subsection{Results and discussions}

We derive constraints on ALPs from the FAST observation for the ALP decay and conversion scenarios, respectively. 
For the ALPs decay,
the 95\% upper limit on $g_{a\gamma}$ has been set within the ALP mass range [8.28, 12.40] $\mu \rm eV$ as shown in the left panel of Fig.~\ref{fig:axion}. 
While for the ALPs conversion, the FAST observation has enabled the exploration of the ALP mass range [4.14, 6.20] $\mu \rm eV$ and the 95\% upper limit is displayed in the right panel of Fig~\ref{fig:axion}. {In both panels, we mask the data from FAST observations in the frequency band $[1160-1300]\, {\rm MHz}$ due to the existence of strong contamination caused by RFI.} Additionally, we discard the data in the frequency ranges $[1000-1005]\, {\rm MHz}$ and $[1465-1500]\, {\rm MHz}$ due to degraded quality. 
From Fig~\ref{fig:axion}, we can clearly notice the presence of some mass points, which exhibit significantly weaker constraints compared with others, such as the ALP mass around $9.0\,\mu$eV ($4.5\,\mu$eV) and $11.74\,\mu$eV ($5.87\,\mu$eV) in the case of ALP decay (conversion). The former corresponds to frequencies near 1090 MHz, where daytime observations are affected by civil aviation\footnote{https://fast.bao.ac.cn/cms/category/rfi\_monitoring\_en/}, resulting in a weaker constraint. The latter corresponds to frequencies near 1420 MHz, where the neutral hydrogen 21-cm emission would be significant. Our constraints, which could serve as a  complimentary check for the non-detection of ALP from ground experiments, for both (decay/conversion) mechanisms are not as strong as the existing limits from CAST~\citep{CAST:2017uph}.

As the largest filled-aperture radio telescope, FAST is designed for a total bandwidth from 70 MHz to 3 GHz~\citep{2020RAA....20...64J}, corresponding to a broader sensitive ALP mass range. The expected {root mean square (RMS)} of the FAST observation in a bandwidth $\mathcal{B}$ is

\begin{equation}
    \sigma_{T} = \frac{T_{\rm sys}}{\sqrt{2\mathcal{B} t_{\rm obs}}},
\end{equation}
{where $T_{\rm sys}$ for the central beam is about 24 K~\citep{2020RAA....20...64J}}, $t_{\rm obs}$ is the observation time, and $\mathcal{B}=max\{\mathcal{B}_{\rm res}, \Delta\nu \}$ with $\mathcal{B}_{\rm res}$ being the frequency resolution.

To assess the prospective capabilities of the FAST, we have calculated the anticipated constraints on the ALP-photon coupling parameter $g_{a\gamma}$ using 100 hours of observation time, for both ALP decay and conversion scenarios. Here we further consider the uncertainty of $D(3')$ calculated with the publicly available code CLUMPY~\citep{Bonnivard:2015pia} from the stellar kinematic data~\citep{Bonnivard:2015xpq}. In Fig.~\ref{fig:forecast} we show the estimated sensitivity for FAST with a frequency range of 70-3000 MHz to ALP decay and conversion in Coma Berenices respectively. At each ALP mass, the sensitivity of the coupling $g_{a\gamma}$ are calculated with the median $D(3')$ (solid lines) and $\pm 95\%$ confidence interval of $D(3')$. {
In the left panel of Fig.~\ref{fig:forecast}, the expected
constraints for decay scenario display an inflection around several $\mu$eV, which arises from the competition between ALP stimulated decay and spontaneous decay.}
It shows that, only in the most optimistic case, the expected constraints from Coma Berenices could be comparable with those from the CAST helioscope~\citep{CAST:2017uph} and the ADMX haloscope~\citep{ADMX:2001dbg,ADMX:2019uok}. {By the way, RFI contamination may vary depending on the time of observation and the local environment. While it is easy to identify in observed real data, incorporating it into forecasts is challenging. Therefore, in this work, we leave gaps in constraints with real observational data but do not consider it in prospective measurements.}

In this work, we adopt the methodology established in Ref.~\citep{Sigl:2017sew} and Ref.~\citep{Caputo:2018ljp} to compute the conversion signals of ALP DM in the presence of the magnetic field. When examining ALP DM conversion in dSphs, the coherence length of the magnetic field within such systems remains uncertain, resulting in a significant ambiguity in the factor $f(m_{a})$. However, due to the typically weak magnetic fields in dSphs, our work could show that it is not advisable to investigate the ALP DM conversion in the magnetic field of dSphs. But other mechanisms may enhance the ALP conversion in dSphs, in principle. Though the star formation has ended very long before in dSphs, they may still host some compact stars, such as neutron stars and white dwarfs. These compact stars may have much stronger magnetic fields, and the resonant conversion of these ALPs into photons near one single compact star has been investigated recently~\citep{Hook:2018iia, Huang:2018lxq, Leroy:2019ghm, Foster:2020pgt, Witte:2021arp, Zhou:2022yxp}. 
On the other hand, recent work shows that axions can be produced in the polar cap region of pulsars and could form a dense ‘axion cloud’ around the star with a very high density of axions~\citep{Noordhuis:2023wid}. Consequently, enhanced signals, for ALPs conversion, could be anticipated in the radio observations of dSphs~\citep{Safdi:2018oeu, Wang:2021hfb, Foster:2022fxn}, and could be searched by current radio telescopes. We leave this aspect of the work for future investigation.




\section{Conclusion}\label{Conclusion}

{ ALPs, as one of the promising candidates for cold dark matter, have attracted widespread attention in recent years. 
The coupling between ALPs and photons allows for the production of two photons through ALP decay, or one photon through ALP conversion in an external magnetic field via the Primakoff effect, providing a solid basis for the search for ALP dark matter.
In this study, we utilized FAST observations in one of the darkest dwarf galaxies, Coma Berenices, known for its high mass-to-light ratios and low radio background, to search for signals of ALP dark matter.}

{As the largest filled-aperture radio telescope, FAST covers the frequency range of 1.0-1.5 $\rm GHz$, enabling the detection of photons resulting from ALP decay within the mass range of [8.28, 12.40] $\mu \rm eV$ or from ALP conversion within the mass range of [4.14, 6.20] $\mu \rm eV$. In essence, FAST exhibits sensitivity to ALPs with masses ranging from several $\mu \rm eV$ to tens of $\mu \rm eV$. 
This mass range is crucial, as ALPs within it can naturally account for the abundance of dark matter.
With the continuum spectrum observed from Coma Berenices, we derived the 95\% upper limits of the ALP-photon coupling parameter $g_{a\gamma}$, as shown in Fig.\ref{fig:axion}. 
However, our constraints on $g_{a\gamma}$ from the FAST observations of Coma Berenices, for both ALP decay and conversion mechanisms, are weaker than those from ground-based experiments such as the CAST helioscope and the ADMX haloscope. While the observation of dSphs by FAST could potentially yield very strong constraints for WIMPs\citep{Guo:2022rqq}, the search for ALPs can only serve as a complementary check on the absence of ALP detection in ground-based experiments.}

{ Moreover, FAST is designed with a total bandwidth ranging from 70 MHz to 3 GHz, corresponding to the ALP mass range of [0.58, 24.80] $\mu \rm eV$ in the case of ALP decay and [0.29, 12.40] $\mu \rm eV$ in the case of ALP conversion, respectively. 
To assess the potential of the ``whole'' FAST, as illustrated in Fig.~\ref{fig:forecast}, we calculated the projected sensitivity of the ALP-photon coupling parameter assuming an observation time of 100 hours. Our results suggest that, even with the outstanding sensitivity of FAST, surpassing current experimental constraints on ALP dark matter through radio observations of dSphs remains a challenging task,
unless potential enhancements of ALP signals by compact stars within dSphs are taken into consideration\footnote{Our preliminary results, which consider compact stars, are comparable to the constraints from CAST~\citep{CAST:2017uph}. We will investigate it in detail in a forthcoming paper.}.
}

\section*{Acknowledgments}
This work made use of the data from FAST (Fivehundred-meter Aperture Spherical radio Telescope). FAST is a Chinese national mega-science facility, operated by the National Astronomical Observatories, Chinese Academy of Sciences. This work is supported by the National Key Research and Development Program of China (No. 2022YFF0503304), the National Natural Science Foundation of China (Nos. 12003069 and 12322302), the Project for Young Scientists in Basic Research 
of Chinese Academy of Sciences (No. YSBR-061), the Chinese Academy of Sciences, and the Entrepreneurship and Innovation Program of Jiangsu Province.

\bibliography{ff}
\bibliographystyle{elsarticle-harv}

\end{document}